\documentclass[aps,prd,showpacs,amsmath,amssymb]{revtex4}

\usepackage{slashed}
\usepackage[dvips]{color}
\usepackage[dvips]{epsfig}
\usepackage{latexsym}
\usepackage{bm}
\usepackage{upgreek}
\usepackage{mathrsfs}
\usepackage{times}
\usepackage{amsthm}
\usepackage{amssymb}
\usepackage{epsfig}
\usepackage{graphicx}
\usepackage{amsmath}
\usepackage{color}

\definecolor{purple}{rgb}{0.8,0,0.6}
\newcommand{\revision}[1]{{#1}}

\begin{document}
\title{On symmetry and topological origin of Weyl particles}

 \author{G.E.~Volovik}
\affiliation{Low Temperature Laboratory, Aalto University, P.O. Box 15100, FI-00076 Aalto, Finland}
\affiliation{ Landau Institute for Theoretical Physics RAS, Kosygina 2, 119334 Moscow, Russia}

\author{M.A.~Zubkov}
\affiliation{ITEP, B.Cheremushkinskaya 25, Moscow, 117259, Russia }
\affiliation{The University of Western Ontario, Department of Applied Mathematics,
 1151 Richmond St. N., London (ON), Canada N6A 5B7 }

\date{\today}

\begin{abstract}
We suggest that the Weil spinors originate from the multi - component fermion fields. Those fields belong to the unusual  theory that, presumably, exists at extremely high energies. In this theory there is no Lorentz symmetry. Moreover, complex numbers are not used in the description of its dynamics. Namely, the one - particle wave functions are real - valued, the functional integral that describes the second - quantised theory does not contain the imaginary unit as well. In the low energy effective theory the two - component Weil spinors appear. Their appearance is related to the
Atiyah-Bott-Shapiro construction and to the expansion of the real
matrix near the topologically protected nodes in three dimensional momentum space. The complex numbers entering ordinary quantum mechanics emerge together with the Weil fermions. \revision{In this pattern gauge fields and gravitational fields appear as certain collective excitations (of the original theory) experienced by the low - energy Weil fermions.}
\end{abstract}
\maketitle

\section{Introduction}

The idea of Majorana that fermions can be represented
in terms of real numbers, together with momentum space topology, may give a hint
why the complex numbers enter quantum mechanics.
The imaginary unit $i=\sqrt{-1}$ is the product of human mind
which is mathematically convenient. However, all the physical
quantities are real, which means that the imaginary unit
cannot enter any physical equation including the wave function.
As is known, Schr\"odinger strongly resisted to introduce
$i=\sqrt{-1}$ into his wave equations (see Yang \cite{Yang}).

The complex numbers can be represented by $2\times 2$ real matrices
\begin{equation}
c=a+ib\rightarrow C=aI+gb =\left(\begin{array}{cc}a&  b\\
-b&a\end{array}\right)~,
\label{-1element}
\end{equation}
where matrices of real and imaginary units are $I$ and $g$
\begin{equation}
I=\left(\begin{array}{cc}1& 0\\
0&1\end{array}\right)~~,~~g=\left(\begin{array}{cc}0& 1\\
-1&0\end{array}\right)~~,~~g^2=-1 ~.
\label{Units}
\end{equation}
The complex conjugation is defined by matrix $q$, which can be chosen as:
\begin{equation}
q=\left(\begin{array}{cc}1&  0\\ 0&-1\end{array}\right)~~,~~ C^*=qC q=aI-gb~~,~~ c^*=a-ib
\,.
\label{ComplexConjugation}
\end{equation}

It is necessary to find out the reason, why such structure entered
the wave function in quantum mechanics, i.e. what is the geometric, symmetry
or topological origin of the algebra of complex numbers in the formalism of
quantum mechanics.
One scheme has been suggested by Adler (see recent paper \cite{Adler2013}
and references therein). It is the so-called trace dynamics for classical
matrix-valued fields, where the quantum field theory is the emergent
thermodynamics, and $i_{\rm eff}$  appears as an anti-self-adjoint operator:
$i_{\rm eff}=-i_{\rm eff}^\dagger$, $i_{\rm eff}^2=-1$.

Here we consider another possible source: the compex numbers
 come from the analog of the Atiyah-Bott-Shapiro construction (ABS) for
fermions
 \cite{Horava2005,VolovikBook},
which are originally expressed via real numbers only.
{In our approach we assume the existence of the underlined theory working at extremely high energies (presumably, of the  order of Plank mass or higher). In this theory there is no Lorentz symmetry. The multi - component fermions are real. Their one - particle dynamics is given by the translation in time of the real - valued $n$ - component wave function. The multicomponent system is described by the functional integral over grassmann variables that does not contain complex numbers.}

{At low enough energies the emergent two - component spinors appear dynamically. Their appearance is related to the particular momentum space topology in the vicinity of the analogue of the Fermi - point and to the particular symmetry (fermion number conservation and the time reversal symmetry). The dynamics of these spinors is governed by the usual functional integral with the action of Weil - spinors.} {The vierbein and the $U(1)$ gauge field also appear dynamically as certain collective excitations of the high - energy fields.}

\section{The theory under construction}
\label{TimeTranslation}

In this section we give the definition of the theory under construction. There are three levels of the given theory. On the first level we deal with the wave function. In coordinate space it is the real - valued $n$ - component vector. On the second level we consider the second quantization using the path integrals. This is the description of many - particle quantum system, or the field system without interaction between the particles. On the third level we consider the quantum field system with the interaction between the particles. We introduce this interaction as a coupling with the emergent gauge field and with the emergent vierbein.

\subsection{The first level. The evolution in time of the one - particle real - valued wave function.}

In this subsection we rely on the one - particle quantum dynamics and operate with the vector of real - valued wave functions $\chi({\bf x},t)$. The important feature of the developed theory is that these wave functions are real - valued while in the conventional quantum mechanics the wave functions are always complex - valued. In  section \ref{SectFuncInt} we give the formulation of the corresponding field theory based on the functional integrals.

 One assumes the invariance of $\chi({\bf x},t)$ under combined time translation, which can be
written in the following general form:
\begin{equation}
\hat U(t_0,{\bf x})\chi({\bf x},t) =\chi({\bf x},t) ~~,~~\hat U(t_0,{\bf
x})= \exp \left[ t_0 \left(\partial_{t} +  A(\nabla) \right)   \right]\,.
\label{TimeTranslationGeneral}
\end{equation}
 We assume that the vector field $\chi({\bf x},t)$ defined in coordinate space is real.
% though in the Fourier analysis of the real vector field
%$\chi({\bf x})$  it is more convinient to use the complex function $\chi({\bf p}) = \sum_{\bf x} e^{i {\bf p} {\bf x}} \chi({\bf x})$ with
%$\chi^*({\bf p}) =\chi(-{\bf p})$. At this stage we would like to avoid the introduction of the imaginary unit in situation, when the use of complex numbers is caused by the advantage of the Fourier analysis. We discuss here the possibble deep reason of the comlexification in quantum mechanics, which is caused by some underlying fundamental symmetry group. That is why here it is assumed that the Fourier comonents are expressed in terms of real functions of sine and cosine. Also in principle the vector ${\bf p}$ can be simply the 3-component internal parameter of the vacuum, which after Fourier transformation gives rise to the emerging coordinate space.
 The time translation is the combined symmetry
operation, which includes conventional translation in time $t \rightarrow t +
t_0$ and the linear transformation
of the vector field.  The generator of the this transformation is the differential operator $A$, which may be represented as a series in powers of $\nabla_i$ with real - valued matrix coefficients. For example, if $A$ is linear in the spacial derivatives, this operator is represented by the product $ B_a \nabla^a$, where $B_a$ are real - valued matrices that satisfy:
\begin{equation}
B_a^{T}=B_a
\,,
\label{SkewSymmetry}
\end{equation}

For the usual four - component Majorana fermions we have (the so - called Majorana representation of Dirac equation):
\begin{equation}
B_1 =  - \sigma_1 \otimes \tau_1, \quad B_2 = \tau_3, \quad B_3 = -\sigma_3 \otimes \tau_1
\end{equation}

In sect. \ref{SectFuncInt} it will be shown, that matrices $B_a$ should be symmetric.

\subsection{The second level. Path integral quantization.}
\label{SectFuncInt}

On the language of functional integral the evolution in time of the field system is given by the correlations of various combinations of the given fields. In our case of $n$ - component real spinor field the partition function is represented as
\begin{equation}
Z = \int D \psi {\rm exp}\Bigl( -\int d t \sum_{{\bf x}} \psi_{\bf x}^T(t) (\partial_t + \hat{A}) \psi_{\bf x}(t) \Bigr)\label{FI}
\end{equation}
while various correlators of the field $\psi$ are given by
\begin{equation}
\langle \psi_{{\bf x}_1}(t_1)  \psi_{{\bf x}_2}(t_2)  ...  \psi_{{\bf x}_2}(t_2) \rangle= \int D \psi {\rm exp}\Bigl(- \int d t \sum_{{\bf x}} \psi_{\bf x}^T(t)  (\partial_t + \hat{A}) \psi_{\bf x}(t) \Bigr) \psi_{{\bf x}_1}(t_1)  \psi_{{\bf x}_2}(t_2)  ...  \psi_{{\bf x}_2}(t_2)\label{correlator}
\end{equation}
Here $\psi$ is the $n$ - component anti - commuting variable. The real nature of the fermions is reflected by the absence of the conjugated set of variables $\bar{\psi}$ and the absence of the imaginary unit in the exponent. The dynamics of the system is completely described by various correlators of the type of Eq. (\ref{correlator}). It is worth mentioning that the complex numbers do not enter the dynamics described by Eq. (\ref{correlator}). It can be easily seen, that if $A$ is linear in the spacial derivatives, and is represented by the product $ B_a \nabla^a$, then $B_a$ should be symmetric. (For the anti - symmetric $B_a$ expression $\sum_{\bf x}\psi_{\bf x}^T  B_a \nabla^a \psi_{\bf x}$ vanishes.) We feel this instructive to give the representation of the partition function of Eq.(\ref{FI}) in terms of the analogues of the energy levels (see Appendix B). It follows from this representation, that the evolution of the real - valued wave function given by Eq. (\ref{TimeTranslationGeneral}) appears in the solution of the equation, that gives the values of "energy" levels.

\subsection{Repulsion of fermion branches $\rightarrow $ the reduced number of fermion species at low energy.}
\label{TopologyZeroes}

{The notion of  energy  in the theory described by operator $\hat A$ may be based on the definition of the values $E_n$ given in Appendix B. Besides, we may introduce the notion of energy scale $\cal E$  as the typical factor in the dependence of various dimensionless physical quantities $q$ on time:  $q \approx f({\cal E} t)$, where $f$ is a certain dimensionless function of dimensionless argument such that its derivatives are of the order of unity. With this definition of energy it can be shown, that at low energies only the minimal number of fermion components effectively contributes the dynamics. Below we make this statement explicit and present the sketch of its proof.}

{Actually, the similar considerations allow to prove the Horava's conjecture presented in \cite{Horava2005}. (According to this conjecture, in particular, any condensed matter theory with  the Fermi - points may be reduced at low energies to the theory described by the two - component Weil spinors.)
We feel this instructive to present the proof of the latter statement. It  is given in Appendix A. }

{We know, that there is the "repulsion" between the energy levels in ordinary quantum mechanics. Similar situation takes place for the spectrum of $\hat A$.
Operator $\hat{A}$ can be written as a $n \times n$ matrix function $A(\nabla)$ of $\nabla$. As it is explained in Appendix
B, in lattice discretization it is given by the skew - symmetric  $Nn\times Nn$ matrix, where $N$ is the total number of the lattice points while $n$ is the number of the components of the spinor $\psi$. As a result there exists the orthogonal $Nn\times Nn$ lattice transformation $\hat \Omega$ that brings matrix $\hat A$ to the block - diagonal form with the $2\times 2$ blocks of the form $E_k \hat \beta = E_k \left(\begin{array}{cc}0 & -1 \\1 & 0 \end{array}\right)$ with some real values $E_k$. In the continuum language matrix $\Omega$ becomes the operator that acts as a $n\times n$ matrix, whose components are the operators acting on the coordinates. There are several branches of the  values of $E_k$. Each branch is parametrized by the $3D$ continuum parameters.
Several branches of spectrum of $E_k$ repulse each other because they are the eigenvalues of the hermitian operator. This repulsion means, that any small perturbation pushes apart the two crossed branches. That's why only the minimal number of branches of its spectrum may cross each other. This minimal number is fixed by the topology of momentum space  (see below, sect. \ref{sectf}).}

{As it was mentioned, there exists the orthogonal operator $\hat \Omega$ (it conserves the norm $\int d^3 x \chi^T_x \chi_x$) such that the operator
\begin{equation}
\tilde{A} = \hat \Omega^T A \hat \Omega\label{omega}
\end{equation}
 is given by the block - diagonal matrix with the elementary $2\times 2$ blocks:
\begin{equation}
\tilde{A} = \left(\begin{array}{cccc}\hat \beta E_1({\cal P}) & 0 & ...& 0 \\0 & \hat \beta E_2({\cal P}) &  ...& 0 \\
... & ... & ... & ...\\
0 & ... & 0 & \hat \beta E_n({\cal P})
\end{array} \right)\label{blockdiag}
\end{equation}
Here we denote by $\cal P$ the three - dimensional vector that parametrizes the branches of spectrum and the basis vector functions that correspond to the given form of $\hat A$. The first $n_{\rm reduced}$ values $E_k$ coincide at ${\cal P} = {\bf p}^{(0)}$. This value is denoted by $E_0 = E_1({\bf p}^{(0)}) = E_2({\bf p}^{(0)})=...$. The first $2n_{\rm reduced}\times 2n_{\rm reduced}$ block $\tilde{A}_{\rm reduced}$ corresponds to the crossed branches.
The remaining block of matrix $\tilde{A}_{\rm massive}$ corresponds to the "massive" branches. The functional integral can be represented as the product of the functional integral over "massive" modes and the integral over $2 n_{\rm reduced}$ reduced fermion components
\begin{equation}
\Psi_{\rm }({\cal P},t) = e^{E_0 \, \hat \beta \, t } \Pi\,  \psi({\cal P},t)
\end{equation}
Here by $\beta$ we denote $2n_{\rm reduced}\times 2n_{\rm reduced}$ matrix $\beta \otimes 1$, while $\Pi$ is the projector to space spanned on the first $2n_{\rm reduced}$ components.
Let us denote the remaining components of $\psi$ by
\begin{equation}
\Theta_{\rm }({\cal P},t) = (1-\Pi)\, \Omega \psi({\cal P},t)
\end{equation}
 We  arrive at
\begin{eqnarray}
Z &=& \int D \Psi D \Theta {\rm exp}\Bigl(- \int d t \sum_{{\cal P}}\Bigl[ {\Psi}_{\cal P}^T(t) e^{ E_0 \, \hat \beta \, t }(\partial_t + \hat{A}_{\rm reduced}({\cal P}))e^{ -E_0 \, \hat \beta \, t } \Psi_{\cal P}(t)
\nonumber\\&&
+  {\Theta}_{\cal P}^T (\partial_t + \hat{A}_{\rm massive}({\cal P}) )\Theta_{\cal P} \Bigr]\Bigr),\label{FI3}
\end{eqnarray}
where ${A}_{\rm massive}({\cal P}) =  (1-\Pi) {A}_{}({\cal P})(1-\Pi^T)$. }

{The exponent in Eq. (\ref{FI3}) contains the following term that corresponds to the contribution of the fermion fields defined in a  vicinity of ${\cal P} = {\bf p}^{(0)}$:
\begin{equation}
{\cal A}_{p^{(0)}} = \int d t \sum_{{\cal P}, k = 1...n_{\rm reduced}} {\Psi}_{k,{\cal P}}^T(t) (\partial_t +  \beta [E_k ({\cal P}) - E_k({\bf p}^{(0)})]) \Psi_{k,{\cal P} }(t), \label{AH}
\end{equation}
We have the analogue of the $2n_{\rm reduced}\times 2n_{\rm reduced}$ hamiltonian $H({\cal P}) = [E_k ({\cal P}) - E_k({\bf p}^{(0)})]$ that vanishes at ${\cal P} = {\bf p}^{(0)}$. Following Appendix B we come to the conclusion, that in the expression for the partition function Eq. (\ref{G_p_h_E_}) the small values of energies  dominate (when the negative energy states are occupied), and these energies correspond to the reduced fermions $\Psi$. It is important, that in order to deal with vacuum, in which negative energy states for Eq. (\ref{AH}) are occupied we need to impose the antiperiodic boundary conditions in time on  $\Psi$ (not on the original fermion field $\psi$).
The other components $\Theta$ contribute the physical quantities with the fast oscillating factors because they are "massive", i.e. do not give rise to the values of $E_n$ from the vicinity of zero. Therefore, these degrees of freedom may be neglected in the description of the long - wavelength dynamics.

Thus the zeroes of $A_{\rm reduced}$ should correspond to the crossing points of the minimal number of branches of the spectrum. They should be described in terms of the exceptional crossing points \cite{NeumannWigner,Novikov1981}. In the following sections, speaking of the low energy dynamics, we shall always imply, that $\hat A_{\rm reduced}$ is discussed, and shall omit the superscript "$\rm reduced $".
The zeros of $\hat A$ should be topologically protected; i.e. they must be robust to deformations.
These should be the point zeroes in 3D space of parameters.
It is worth mentioning, that for these point zeros to dominate there should exist the underlying discrete or continuous symmetry which forbids the existence of the more powerful manifolds of zeroes -- Fermi surfaces and nodal lines in the spectrum.

\revision{Any basis of the wave functions is related via an orthogonal operator $\hat{\tilde{\Omega}}$ to the basis of the wave functions, in which $\hat{A}_{\rm reduced}$ has the form of the block - diagonal matrix (Eq. (\ref{blockdiag})). We require, that $\hat{\tilde{\Omega}}$ commutes with $\hat \beta$ for the transformation to the basis associated with the observed low energy coordinates. This observed coordinate space may differ from the primary one, so that  $\hat{\tilde{\Omega}}$ is not equal to $\hat{{\Omega}}$ of Eq. (\ref{omega}). This means, that the coordinate space in not the primary notion, but rather the secondary one. $[\hat{\tilde{\Omega}},\beta]=0$ is the requirement, imposed on the representation of the theory, that allows to recover the usual Weil spinors and the conventional quantum mechanics with complex - valued wave functions (see the next subsection).
In this basis $\hat A_{\rm reduced}$ is given by the differential operator. It is expressed as a series in powers of derivatives with real - valued $2n_{\rm reduced}\times 2n_{\rm reduced}$ matrices as coefficients. From $[\hat{\tilde{\Omega}},\beta]=0$ it follows, that in this basis $[\hat A_{\rm reduced} , \hat \beta] = 0$.}

\subsection{Fermion number conservation for the $4$ - component spinors}

{In Section \ref{TopologyZeroes} it was argued that the number of fermion components at low energies should be even. The minimal even number that admits nontrivial momentum space topology (see below) is $4$.
That's why we consider the effective low energy four - component spinors. \revision{Besides, it was argued, that for the low energy effective fermion fields the reduced operator $\hat A_{\rm reduced}$ in a certain basis commutes with $\beta = \left(\begin{array}{cc}0 & -1\\1 & 0\end{array} \right)$.}
\revision{This may be identified with the fermion number conservation, that is rather restrictive. Below we describe the general form of the $4\times 4$ operator $\hat A$, and demonstrate how this requirement reduces its general form.} Let us introduce the two commuting momentum operators:
\begin{equation}
\hat {\cal P}_{\beta} = - \hat{\beta} \, \nabla, \quad \hat {\cal P}_{\alpha} = - \hat{\alpha} \, \nabla
\end{equation}
where
\begin{eqnarray}
&&  \hat \beta = -1\otimes \hat \tau_3 \hat \tau_1 = -1\otimes i \tau_2,~~ \hat \alpha = -\hat \sigma_3 \sigma_1 \otimes 1 = - i \sigma_2 \otimes 1
\end{eqnarray}}

{The two commuting operators $\hat {\cal P}_{\beta}$ and $\hat {\cal P}_{\alpha}$ have common real - valued eigenvectors corresponding to their real - valued eigenvalues.}
{Matrix $A$ can be represented as the analytical function
\begin{eqnarray}
\hat A &=&  {\cal F} (\hat {\cal P}_{\beta}, \hat {\cal P}_{\alpha}, \hat L^k, \hat S^k ),
\end{eqnarray}
where
\begin{eqnarray}
&&  \hat L^k=(\hat \sigma_1 \otimes \hat \beta, -\hat \alpha \otimes 1,  \hat \sigma_3 \otimes \hat \beta),\nonumber \\
&&  \hat S^k=(\hat \alpha \otimes \hat \tau_1, -\hat 1\otimes \beta,  \hat \alpha \otimes \hat \tau_3 ),\nonumber \\
\label{10matrices}
\end{eqnarray}
More specifically, it can be represented as
\begin{eqnarray}
\hat A &=&   \sum_{k=1,2,3} m^{L}_k( {\cal P}_{\beta})\hat L^k  + \sum_{k=1,2,3}m^{S}_k( {\cal P}_{\alpha})\hat S^k \nonumber\\ && +  m^I_1({\cal P}_{\beta}) \hat I^1 - m^I_2({\cal P}_{\alpha}) \hat I^1 + m^I_3({\cal P}_{\beta}) \hat I^3 - m^I_4({\cal P}_{\alpha}) \hat I^3  + m^o({\cal P}_{\beta}) \hat \beta\,,
\label{Ageneral}
\end{eqnarray}
Here
\begin{eqnarray}
&& \hat I^1 = \hat \sigma_1 \otimes \hat \tau_3, \quad \hat I^3 = \hat \sigma_3 \otimes \tau_3
\end{eqnarray}
while  $m^{L}_k({\cal P}), m^{S}_k({\cal P}),m^I_k({\cal P}), m^o({\cal P})$ are real - valued  functions of the momenta ${\cal P}$. Functions $m^{I}_k({\cal P}), m^{o}_k({\cal P})$ are odd;  $\hat L^k$ and $ \hat S^k$ are the generators of the two $SO(3)$ groups; $\hat \beta$ and $\hat \alpha$ are real antisymmetric matrices that commute with all $\hat L^k$ (or $\hat S^k$) correspondingly; $\hat I^k$ are the matrices that commute  with $\hat \alpha \otimes \hat \beta$ but do not commute with either of $\hat \alpha$ and $\hat \beta$.
(Notice, that $\hat \beta \hat L^2 =  \hat \alpha \hat S^2 = -\alpha \otimes \beta$. That's why odd part of the function $m^{S}_2$ may be set equal to zero. )}

\revision{It has been explained at the end of Section \ref{TopologyZeroes}, that for the operator $\hat A$ there always exists the representation, in which it commutes with matrix
$ \hat \beta = 1 \otimes (-i \tau_2)$. We assume, that this is the representation associated with the observed coordinate space.} Matrix $\hat \beta$ anticommutes with $ \hat{S}_k$, $k=1,3$ and $\hat{I}_k$, $k=1,2,3,4$. {Yet another way to look at this symmetry is to require, that the momentum defined as
$\hat {\cal P}_{\beta} = - \hat \beta \nabla$ is conserved, i.e. commutes with $\hat A$.} This requirement reduces the partition function to
\begin{equation}
Z = \int D \psi {\rm exp}\Bigl(  -\int d t \sum_{{\bf x}} {\psi}^T_{{\bf x}}(t) (\partial_t + i_{\rm eff} m^L_k(\hat {\cal P}_{\beta}) \hat \Sigma^k  + i_{\rm eff} m(\hat {\cal P}_{\beta}))\psi_{\bf x}(t) \Bigr)\label{Z__0}
\end{equation}
where we introduced the effective $4\times 4$ imaginary unit
\begin{equation}
i_{\rm eff}=\hat \beta\,,
\label{effective_i}
\end{equation}
Thus operator $\hat A$ can be represented as the analytical function of ${\cal P}_{\beta}$ and $\hat L^k$ only:
$
\hat A =  {\cal F} (\hat {\cal P}_{\beta},  \hat L^k)
$.
Here we have introduced the $4\times 4$ matrices forming the quaternion units $\Sigma_k$, that can be
represented in terms of the $2\times 2$ complex Pauli matrices.
Matrix $1\otimes \tau_3$ becomes the operator of complex conjugation in Eq.(\ref{ComplexConjugation}).

The three real - valued $4\times 4$ $\Sigma$-matrices form the basis of the $su(2)$ algebra and have the representation in terms of the three complex Pauli matrices:
\begin{eqnarray}
\Sigma^1= \sigma^1 \otimes 1~,~  \Sigma^2= i_{\rm eff} \Sigma^1 \Sigma^3~,~
\Sigma^3=\sigma^3 \otimes 1\,
\label{sigma_matrices}
\label{gamma-sigma}
\end{eqnarray}

\subsection{Time reversal symmetry}
%\item{}

We impose the time reversal (or, CP) symmetry generated by ${\cal T} = - i \sigma^2 \tau^3 = \hat S^3$ and followed by the change ${\bf x} \rightarrow -{\bf x}$. Its action on the spinors is:
\begin{eqnarray}
&& {\cal T} \psi({\bf x}) = - i \sigma^2 \tau^3 \psi(-{\bf x})
\end{eqnarray}
It prohibits the term with $m({\cal P})$. Thus operator $\hat A$ can be represented as
\begin{eqnarray}
\hat A &=&  {\cal F} (\hat {\cal P}_{\beta},  \hat L^k) =  \sum_{k=1,2,3} m^{L}_k( {\cal P}_{\beta})\hat L^k
\end{eqnarray}

\subsection{Topology of zeroes}
\label{sectf}

The topologically nontrivial situation arises when
${\bf m}^{L}({\cal P})$ has the hedgehog singularity.
The hedgehog point zero is described by the topological invariant
\begin{equation}
N= \frac{e_{ijk}}{8\pi} ~
   \int_{\sigma}    dS^i
~\hat{\bf m}\cdot \left(\frac{\partial \hat{\bf m}}{\partial p_j}
\times \frac{\partial \hat{\bf m}}{\partial p_k} \right) \,,
\label{N}
\end{equation}
where $\sigma$  is the $S^2$ surface around the point.

For the topological invariant $N=1$ in Eq.(\ref{N}) the expansion near the hedgehog point at $P^{(0)}_j$ in $3D$ ${\cal P}$-space gives
\begin{equation}
m_i({\cal P})=f_i^j({\cal P}_j-P^{(0)}_j)\,.
\label{A(K)-expansion0}
\end{equation}

{As a result, Eq. (\ref{Z}) has the form:
\begin{equation}
Z = \int D \psi {\rm exp}\Bigl(-  \int d t \sum_{{\bf x}} {\psi}^T_{{\bf x}}(t) (\partial_t + i_{\rm eff} f_k^j(\hat{\cal P}_j-P^{(0)}_j) \hat \Sigma^k )\psi_{\bf x}(t) \Bigr)\label{Z}
\end{equation}}

{Operator $\hat A$ can be written in the basis of the eigenvectors of $\cal P$. As a result, we arrive at the $4\times 4$ matrix function of real variable $\cal P$:
$A({\cal P})= i_{\rm eff} m^L_k( {\cal P}) \hat \Sigma^k$.
The matrix $A({\cal P})$ near the node has the form
\begin{equation}
A({\cal P})=  i_{\rm eff} \Sigma^i f_i^j({\cal P}_j-{\cal P}^{(0)}_j)\,.
\label{A(K)sigma}
\end{equation}}

This is in some sense similar to the result of \cite{Horava2005} for the low energy effective action for the fermionic condensed matter models and corresponds to the ABS construction of the K - theory applied to momentum space topology.

The topological invariant responsible for the singularity can be written
analytically if one considers the extended matrices $A(P_{\mu})\equiv A({\cal
P},{P_4})=P_4+ A({\cal P})$, where $P_4$ are the eigen values of the operator
$\partial_t$. As a result, the $\pi_2$ homotopy group is extended to $\pi_3$
(compare with the generator of $\pi_3(R_n)$ for $n>3$ on page 133 of Ref.
\cite{Whitehead1942}):
\begin{equation}
N = \frac{e_{\alpha\beta\mu\nu}}{48\pi^2}~
{\bf Tr}\int_\sigma   dS^\alpha
~ A^{-1} \partial_{p_\beta} A~ A^{-1}  \partial_{p_\mu}  A ~A^{-1}
\partial_{p_\nu}  A\,.
\label{TopInvariantA}
\end{equation}
Here $\sigma$ is the $S^3$ spherical surface around the node in 4D
$p_\mu$-space.
%It is possible to bring
%every skew-symmetric matrix to a block diagonal form by an orthogonal
%transformation. Specifically, every $2n \times 2n$ real skew-symmetric
%matrix can be written in the form $A = Q A_0 Q^T$,  where $Q$ is orthogonal
%matrix and $A_0$ is block diagonal.
%Thus the space of the skew-symmetric matrices is ${\cal
%R}=SO(2n)/U(n)$. It has the nonzero homotopy group
%$\pi_2({\cal R})=Z$.
%For the condition (iii) for denenerate points of avoiding crossing
%the relevant space is  ${\cal R}=SO(2n)/(SO(2))^n$. It also has the nonzero homotopy group
%$\pi_2({\cal R})=Z^{n}$, which is consistent with the previous homotopy.

%The contribution of $\hat S^k$ is prohibited, if we impose the following symmetry condition:
%the matrices $A$ must commute with two of the $\hat S^k$ matrices, for example with $\hat S^3=\hat\nu_3 \hat g_\mu$ and $\hat %S^2= \hat g_\nu$. Then all three matrices $\hat S^k$  are prohibited.
%This corresponds to two discrete symmetries, which we discuss later,  with the matrix $\hat g_\nu$ playing the role of imaginary unit.

%This procedure recalls the discussion of the $\pi_3$ homotopy group in  \cite{Whitehead1942} for the orthogonal
%matrices. The $\pi_3$ homotopy group of $2n \times 2n$ matrix is effectively reduced
%to $\pi_3$ of real $4\times 4$ matrices, which can be represented in terms of $\hat L^k$. We shall later return to this connection.

%\section{The origin of Weyl fermions and of the imaginary unit}

\subsection{Propagator, Hamiltonian and Schr\"odinger equation}
\label{PropagatorHamiltonian}

It follows from the functional integral representation, that one can
introduce the propagator (the Green's function) and the Hamiltonian:
\begin{equation}
G^{-1}= i_{\rm eff} A(P_\mu)\equiv - H_{\cal P} +  i_{\rm eff}P_4\,.
\label{GreenFunction}
\end{equation}
This means that the Green's function here is determined on the imaginary
axis.

In terms of the Green's function the topological invariant $N$ in
Eq.(\ref{TopInvariantA}) has the following form:
\begin{equation}
N = \frac{e_{\alpha\beta\mu\nu}}{48\pi^2}~
{\bf Tr}\int_\sigma   dS^\alpha
~ G\partial_{p_\beta} G^{-1}
G\partial_{p_\mu} G^{-1} G\partial_{p_\nu}  G^{-1}\,,
\label{MasslessTopInvariant3D}
\end{equation}
where $\sigma$ is $S^3$ surface around the Fermi point in
$(P_1,P_2,P_3,P_4)$ space.

If the Hamiltonian belongs to topological class $N=1$ or $N=-1$, it
can be adiabatically deformed to the Weyl Hamiltonian for the right-handed
and left-handed fermions respectively:
 \begin{equation}
H=N\left(\Sigma^1 P_x + \Sigma^2 P_y + \Sigma^3 P_z\right) ~~, ~~N= \pm 1
\,,
\label{Weyl}
\end{equation}
where the emergent Pauli matrices $\Sigma^i$ describe the emergent
relativistic spin.
Weyl fermions are the "primary" objects, which emerge in the low-energy
corner. The other ingredient of the Standard Model is the gauge fields. The $U(1)$ gauge field appears as the fluctuations of the fermi - point $p^{(0)}$. The fluctuations of different fermi - points that may depend on various spinor fields give rise to the nonabelian gauge fields \cite{VolovikBook}.

From Eq.(\ref{TimeTranslationGeneral}) it is seen that the matrix $i_{\rm
eff}$,  which commutes with the Hamiltonian, corresponds to the imaginary unit
in the time dependent Schr\"odinger equation.
The latter is obtained, when $p_4$  is substituted by the operator of time
translation,
$p_4\rightarrow \partial_t$:
\begin{equation}
i_{\rm eff} \partial_t \chi=  H\chi\,.
\label{Schroedinger}
\end{equation}

{The whole wave dynamics may be formulated in terms of real functions only. The Hamiltonian is expressed through the momentum operator $\hat{\cal P}_{\beta}$. Its eigenvalues are parametrized by the eigenvalues $\cal P$ of momentum, the projection $n = \pm 1$ of emergent spin $\hat \Sigma$ on vector $m({\cal P})$, and the eigenvalue $C = \pm 1$ of the conjugation operator  $\hat{C} = 1\otimes \tau_3$:
\begin{equation}
\mid  C, n,  {\cal P}\rangle \equiv \Bigl[e^{\frac{1}{2} i_{\rm eff} \hat{\Sigma} \phi[m({\cal P})]}\times e^{i_{\rm eff} {\cal P} {\bf x}}\Bigr] \mid C \rangle \otimes \mid n \rangle,
\end{equation}
where  $\mid  C \rangle = \frac{1}{2}\left(\begin{array}{c} 1 + C\\ -1 + C  \end{array}\right)$, and $\mid  n \rangle = \frac{1}{2}\left(\begin{array}{c} 1 + n\\ -1 + n  \end{array}\right)$,
while  rotation around the vector $\phi$ by the angle equal to its absolute value transforms a unit vector directed along the third axis into the one directed along $m({\cal P})$. Vectors $\mid  C, n,  {\cal P}\rangle$ are the eigenvectors of Hamiltonian correspondent to the eigenvalues $E = C |m({\cal P})|$. Once at $t = 0$ the wave function is given by $\mid  C, n,  {\cal P}\rangle$, its dependence on time is given by:
\begin{equation}
\chi(t) = e^{-i_{\rm eff} C |m({\cal P})|t} \mid  C, n,  {\cal P}\rangle
\end{equation}}

\subsection{The third level of the theory. Interaction between the fermions. Emergent gauge field and emergent vierbein.}
\label{sectthird}

At extremely high energies the partition function for the fermions with the interaction between them can be written in the form:
\begin{equation}
Z = \int D \psi D\Phi {\rm exp}\Bigl(-R[\Phi] - \int d t \sum_{{\bf x}} \psi_{\bf x}^T(t) (\partial_t + \hat{A}(\Phi)) \psi_{\bf x}(t) \Bigr)\label{FIn}
\end{equation}
Here the new fields that provide the interaction between the fermions are denoted by $\Phi$. $R$ is some function of these fields.  Now matrix $\hat A$ also depends on these fields. In mean field approximation, when the values of $\Phi$ are set to their "mean" values we come back to the consideration of the previous subsections. However, at the end of the consideration the fluctuations of the fields $\Phi$ are to be taken into account via the fluctuations of the emergent vierbein $e^a_k$ and the Fermi - point $p^{(0)}$.

{Let us consider first for the simplicity the low energy effective theory with only one emergent Weil fermion. The interaction between the particles appears when the fluctuations of $p^{(0)}_k$ and $f^j_k$ are taken into account. {First, we assume, that these fluctuations are long - wave, so that the corresponding variables should be considered as if they would not depend on coordinates.} We suppose, that when the fluctuations of $\Phi$ are taken into account, the fermion number is conserved. As for, the time reversal symmetry, the fluctuations of the fields $\Phi$ break it.  As a result the partition function of the theory receives the form
\begin{eqnarray}
Z &=& \int D \psi D\Phi {\rm exp}\Bigl( - \int d t \sum_{{\bf x}} {\psi}^T_{{\bf x}}(t) (\partial_t + i_{\rm eff} m^L_{\Phi, k}(\hat {\cal P}) \hat \Sigma^k  + i_{\rm eff} m_{\Phi}(\hat {\cal P}))\psi_{\bf x}(t) \Bigr)
\label{Z__}
\end{eqnarray}}

{Here
\begin{equation}
m^{L}_{\Phi, i}({\cal P})\approx e \, e_i^j({\cal P}_j + B_j ),  \quad m_{\Phi}({\cal P}) \approx B_0 + e\, e_0^j({\cal P}_j+B_j ),\quad i,j = 1,2,3
\label{A(K)-expansion012}
\end{equation}
We represented here $f_i^j = e \, e_i^j$, where the fluctuating long - wave fields $e[\Phi],B[\Phi]$ depend on the primary fields $\Phi$. This representation for $f_i^j$ is chosen in this way in order to interpret the field $e_i^j$ as the vierbein. We require
$e^0_a = 0$ for $a=1,2,3$, and $e \times e_0^0=1$. Here $e^{-1} = e_0^0 \times {\rm det}_{3\times 3}\, e^i_a = e_0^0$ is equal to the determinant of the vierbein $e^i_a$. In the mean field approximation, $\Phi$ is set to its mean value $\Phi_0$, while $B_0[\Phi_0] = e^k_0[\Phi_0] = 0$, and $e\, e^k_a[\Phi_0] = f^k_a$, where  variable $f$ was introduced in sect. \ref{sectf}.}

Besides, we define the new two component spinors starting from the four - component spinor $
\psi = \Bigl(\psi^1,  \psi^2,  \psi^3,  \psi^4 \Bigr)^T$.
Those two - component spinors are given by
\begin{equation}
\Psi({\bf x}) = \left(\begin{array}{c}\psi^1({\bf x}) + i \psi^2({\bf x}) \\ \psi^3({\bf x}) + i \psi^4({\bf x}) \end{array} \right), \quad
\bar{\Psi}({\bf x}) = \left(\begin{array}{c}\psi^1({\bf x}) - i \psi^2({\bf x}) \\ \psi^3({\bf x}) - i \psi^4({\bf x}) \end{array} \right)^T\label{barPsi}
\end{equation}

{As a result, the partition function of the model may be rewritten as:
\begin{equation}
Z = \int D \Psi D\bar{\Psi} D e^i_k D B_k e^{i S[e^j_a, B_j,  \bar{\Psi},\Psi]}
\end{equation}
with
\begin{eqnarray}
S &=&   S_0[e, B] + \frac{1}{2} \Bigl(\int d t \,e\,  \sum_{{\bf x}} \bar{\Psi}_{{\bf x}}(t)  e_a^j  \hat \sigma^a  \hat D_j  \Psi_{\bf x}(t) + (h.c.)\Bigr),\label{Se}
\end{eqnarray}
where the sum is over $a,j = 0,1,2,3$ while $\sigma^0 \equiv 1$, and $\hat D$ is the covariant derivative that includes the $U(1)$ gauge field $B$. $S_0[e, B]$ is the part of the effective action that depends on $e$ and $B$ only.}

{Recall, that we have considered the long - wavelength fluctuations of the emergent fields $B$ and $e$. That is we neglected the derivatives of these fields. In the fermion part of the action in Eq. (\ref{Se}) there are no dimensional parameters.
The only modification of this action that is analytical in $B, e$ and their derivatives and that does not contain the dimensional parameters is if the covariant derivative $D$ receives the contribution proportional to the derivative of $e$. That's why, even for the  non - homogenious variations of $e$ and $B$ in low energy approximation we are left with effective action of the form of Eq. (\ref{Se}) { if the value of the emergent electromagnetic field is much larger than the order of magnitude of quantity $|\nabla e^k_a|$}. Such a situation takes place, for example for the consideration of the emergent gravity in graphene \cite{VZ2013gr}.}

\revision{One can see, that if $S_0[e, B]$ may be neglected, only the second term of Eq. (\ref{Se}) contributes the dynamics, the fields $e^{\mu}_k$ and $B_{\mu}$ may be identified with the true gravitational field (vierbein) and the true gauge field correspondingly. Their effective action is obtained as a result of the integration over the fermions.}

\section{Conclusions}
\label{Discussion}

The complexification of quantum mechanics may be the emergent phenomenon,
which arises in the low energy corner of a system, which experiences the
time translation symmetry.  The latter is the combined symmetry of a system
under transformation, in which the ordinary  time shift
$t\rightarrow t+t_0$ is accompanied by the  linear transformation of the multi - component
real vector functon $\chi$, which describes the state of the
system.
\revision{The generators of this linear transformation
are the differential operators that may be represented as the series in powers of the derivatives with the real matrices as coefficients. The generator $\hat A$ of the time translation enters the formulation of the theory through the functional integrals. In this formulation we have the only set of the anticommuting variables $\psi$. There is no the second set $\bar{\psi}$ that would enter the functional integral for usual spinor fields. Complex numbers do not enter the functional integral as well. In a certain basis of the wave functions the operator $\hat A$ becomes the skew - symmetric block - diagonal matrix with the $2\times2$ blocks. In the latter representation (due to the repulsion of the branches of the spectrum of operator $\hat A$) at low energies only the minimal number of components of the original spinors dominate the dynamics. Nontrivial momentum space topology fixes this minimal number equal to four. There always exists the basis of the fermion wave functions, in which the reduced $4\times 4$ operator $\hat A$ commutes with $\beta = - i\tau_2 \otimes 1$ (corresponds to the fermion number conservation). This basis is associated with the observed coordinate space. In addition, we assume, that at low energies the time reversal symmetry takes place. These two symmetries (fermion number conservation and time reversal symmetry) provide, that in momentum space in the vicinity of the topologically protected point zeros the system is effectively described by real $4\times 4$ matrix, which assembles the
complex structure \cite{StoneChiuRoy2011} in terms of the $2\times 2$ Pauli
matrices.} This gives rise to the effective
Hamiltonian, which describes Weyl fermions. They serve as  primary objects
for the Standard Model, while the other objects are composite and they
inherit the structure of the Weil quantum mechanics.
The interaction between the multi - component fermions of the underlined high - energy system causes the appearance of the emergent gravity in the low energy effective model with one Weil spinor. This quantum gravity corresponds to the fluctuating vierbein. Besides, we have the fluctuating $U(1)$ gauge field.

Since the complex structure of quantum mechanics is the low-energy emergent
phenomenon, it will be lost at sufficiently high energy (this complex-real
border energy scale could be of the order of or above the Planck scale).
All consideration could be presented in momentum space only, so that the coordinate space is simply the emergent phenomenon, which follows from the matrix structure in momentum space
\cite{CortesSmolin2013}. Thus, the coordinate space emerges together with Weyl fermions,
gravity and gauge fields.

The work of M.A.Z. is  supported by the Natural Sciences and Engineering Research Council of
Canada. GEV thanks Yu.G. Makhlin for fruitful discussions and
acknowledges a financial support of the Academy of Finland and its COE
program.

\subsection*{Appendix A. The Horava's conjecture}
\label{horava}

{Here we consider the fermionic theory of $n$ - component complex spinors $\psi$. The partition function has the form:
\begin{equation}
Z = \int D \psi D \bar{\psi} {\rm exp}\Bigl(i\int d t \sum_{{\bf x}} \bar{\psi}_{\bf x}(t) (i\partial_t - \hat{H}_{\rm }) \psi_{\bf x}(t) \Bigr),\label{ZH}
\end{equation}
Let us suppose, that the Hamiltonian $H$ is the hermitian matrix function of momentum $\hat{\bf p} = - i \nabla$. First, we consider  the particular case, when the coefficients in the expansion of $H$ in powers of $\bf p$ do not depend on coordinates.
We know, that there is the "repulsion" between the energy levels in ordinary quantum mechanics. Similar situation takes place for the spectrum of $\hat H$.
The eigenvalues of $\hat H$ are the real - valued functions of $\bf p$.}

{Several branches of spectrum for the  hermitian operator $\hat H$ repulse each other, i.e. any small perturbation pushes apart the two crossed branches. That's why only the minimal number of branches of its spectrum may cross each other. This minimal number is fixed by the topology of momentum space that is the space of parameters $\bf p$.}

{Let us consider the position ${\bf p}^{(0)}$ of the crossing of $n_{\rm reduced}$ branches of $\hat H$. There exists the Hermitian matrix $\Omega$ such that the matrix $\tilde{H}( {\bf p}) = \Omega^+ \hat{H} \Omega$ is diagonal. In this matrix the first $n_{\rm reduced}\times n_{\rm reduced}$  block $\hat{H}_{\rm reduced}$ corresponds to the crossed branches (i.e. all  eigenvalues of $\hat{H}_{\rm reduced}({\bf p})$ coincide at ${\bf p} = {\bf p}^{(0)}$). The remaining block of matrix $\hat{H}_{\rm massive}$ corresponds to the "massive" branches. The functional integral can be represented as the product of the functional integral over "massive" modes and the integral over $n_{\rm reduced}$ reduced fermion components
\begin{equation}
\Psi_{\rm }({\bf x}) = \Pi\, \Omega \psi({\bf x}), \quad \bar{\Psi}_{\rm }({\bf x}) = \bar{\psi}({\bf x})\Omega^+ \Pi^+
\end{equation}
Here $\Pi$ is the projector to space spanned on the first $n_{\rm reduced}$ components. }
{Let us denote the remaining components of $\psi$ by
\begin{equation}
\Theta_{\rm }({\bf x}) = (1-\Pi)\, \Omega \psi({\bf x}), \quad \bar{\Theta}({\bf x}) = \bar{\psi}({\bf x})\Omega^+(1-\Pi^+)
\end{equation}}

{Let us denote the only eigenvalue of $\hat{H}_{\rm reduced}({\bf p}^{(0)})$ by $E_0$. The transformation
$\psi_{\bf x} \rightarrow e^{-i E_0 t} \psi_{\bf x}$ , $H({\bf p}) \rightarrow H({\bf p}) - E_0$ leaves the expression in exponent of Eq. (\ref{ZH}) unchanged.  That's why we can always consider the matrix ${H}_{\rm reduced}$ equal to zero at the position of the branches crossing ${\bf p}^{(0)}$. We are left with the following expression for the partition function:
\begin{equation}
Z = \int D \Psi D\bar{\Psi} D \Theta D\bar{\Theta} {\rm exp}\Bigl(i\int d t \sum_{{\bf x}} \Bigl[\bar{\Psi}_{\bf x}(t) (i\partial_t - \hat{H}_{\rm reduced}) \Psi_{\bf x}(t) + \bar{\Theta}_{\bf x}(t) (i\partial_t - \hat{H}_{\rm massive}) \Theta_{\bf x}(t)\Bigr]\Bigr)\label{ZH2}
\end{equation}
where ${H}_{\rm massive} = (1-\Pi) \hat{H}  (1-\Pi^+)$.       }

{Spectrum of operator $\hat H_{\rm reduced}$ has exceptional properties around vanishing eigenvalues. The corresponding eigenfunctions do not depend on time.   The key point is that at low energy scales the integral over $\Psi_{\rm }({\bf x})$ dominates. The other components $\Theta$ contribute the physical quantities with the fast oscillating factors, and, therefore, may be neglected in the description of the long - wavelength dynamics. As a result at the low energies we may  deal with the theory that has the following partition function:
\begin{equation}
Z = \int D \Psi D\bar{\Psi}  {\rm exp}\Bigl(i\int d t \sum_{{\bf x}} \bar{\Psi}_{\bf x}(t) (i\partial_t - \hat{H}_{\rm reduced}) \Psi_{\bf x}(t) \Bigr)\label{ZH3}
\end{equation}}

{This consideration allows to prove the Horava's conjecture presented in \cite{Horava2005}. According to this conjecture any condensed matter theory with fermions and with the topologically protected Fermi - points may be reduced at low energies to the theory described by the two - component Weil spinors. The remaining part of the proof is the consideration of momentum space topology. It protects the zeros of $\hat H_{\rm reduced}$ (i.e. it is robust to deformations) only when there is the corresponding nontrivial invariant. The minimal number of fermion components that admits nontrivial topology is two.}

This reduces the partition function to
\begin{equation}
Z = \int D \Psi D \bar{\Psi} {\rm exp}\Bigl(  i \int d t \sum_{{\bf x}} \bar{\Psi}_{{\bf x}}(t) (i\partial_t - m^L_k(\hat{\bf p}) \hat \sigma^k  - m(\hat{\bf p}))\Psi_{\bf x}(t) \Bigr)\label{ZH__0}
\end{equation}
where we functions $m^L_k, m$ are real - valued.

Let us, in addition,  impose the time reversal (or, CP) symmetry generated by ${\cal T} = - i \sigma^2 $ and followed by the change ${\bf x} \rightarrow -{\bf x}$. Its action on the spinors is:
\begin{eqnarray}
&& {\cal T} \Psi({\bf x}) = - i \sigma^2 \bar{\Psi}^T(-{\bf x})
\end{eqnarray}
It prohibits the term with $m({\bf p})$. Thus operator $\hat H$ can be represented as
\begin{eqnarray}
\hat H &=&   \sum_{k=1,2,3} m^{L}_k( {\bf p})\hat \sigma^k
\end{eqnarray}

The topologically nontrivial situation arises when
${\bf m}^{L}({\bf p})$ has the hedgehog singularity.
The hedgehog point zero is described by the topological invariant
\begin{equation}
N= \frac{e_{ijk}}{8\pi} ~
   \int_{\sigma}    dS^i
~\hat{\bf m}\cdot \left(\frac{\partial \hat{\bf m}}{\partial p_j}
\times \frac{\partial \hat{\bf m}}{\partial p_k} \right) \,,
\label{NH}
\end{equation}
where $\sigma$  is the $S^2$ surface around the point.

For the topological invariant $N=1$ in Eq.(\ref{NH}) the expansion near the hedgehog point at $p^{(0)}_j$ in $3D$ ${\bf p}$-space gives
\begin{equation}
m_i({\bf p})=f_i^j({\bf p}_j-p^{(0)}_j)\,.
\label{A(K)-expansion0H}
\end{equation}

{As a result, Eq. (\ref{ZH__0}) has the form:
\begin{equation}
Z = \int D \Psi D\bar{\Psi} {\rm exp}\Bigl(i\int d t \sum_{{\bf x}} \bar{\Psi}_{{\bf x}}(t) (i\partial_t - f_k^j(\hat{\bf p}_j-p^{(0)}_j) \hat \sigma^k )\Psi_{\bf x}(t) \Bigr)\label{ZHH}
\end{equation}}

Next, we should consider the situation, when the coefficients of expansion of $H$ in powers of $\bf p$, depend on coordinates and fluctuate. The consideration of this case exactly repeats the one of section \ref{sectthird}. Instead of Eq. (\ref{ZHH}) we obtain:
\begin{equation}
Z = \int D \Psi D\bar{\Psi} D e^i_k D B_k e^{i S[e^j_a, B_j,  \bar{\Psi},\Psi]}
\end{equation}
with
\begin{eqnarray}
S &=&   S_0[e, B] + \frac{1}{2} \Bigl(\int d t \,e\,  \sum_{{\bf x}} \bar{\Psi}_{{\bf x}}(t)  e_a^j  \hat \sigma^a  \hat D_j  \Psi_{\bf x}(t) + (h.c.)\Bigr),\label{SeH}
\end{eqnarray}
where the sum is over $a,j = 0,1,2,3$ while $\sigma^0 \equiv 1$, and $\hat D$ is the covariant derivative that includes the $U(1)$ gauge field $B$. $S_0[e, B]$ is the part of the effective action that depends on $e$ and $B$ only. We represented here by the field $f$ as  $f_i^j = e \, e_i^j$, where the fluctuating long - wave fields are denoted  $e_i^j,B$.  This representation for $f_i^j$ is chosen in this way in order to interpret the field $e_i^j$ as the vierbein. This is achieved by the requirement $e^0_a = 0$ for $a=1,2,3$, and $e \times e_0^0=1$. Here $e^{-1} = e_0^0 \times {\rm det}_{3\times 3}\, e^i_a = e_0^0$ is equal to the determinant of the vierbein $e^i_a$.

\section*{Appendix B. Determinant for real fermions}

{We consider the functional integral over real fermions basing on the analogy with the integral over complex fermions (see \cite{PhysRevD.12.2443}).
We start from the partition function of Eq. (\ref{FI}).
In lattice discretization the differential operator $\hat A$ is represented as the skew - symmetric  $Nn\times Nn$ matrix, where $N$ is the total number of the lattice points while $n$ is the number of the components of the spinor $\psi$. As a result there exists the orthogonal $Nn\times Nn$ transformation $\Omega$ that brings matrix $\hat A$ to the block - diagonal form with the $2\times 2$ blocks of the form
\begin{equation}
E_k \hat \beta = E_k \left(\begin{array}{cc}0 & -1 \\1 & 0 \end{array}\right)
\end{equation}
with some real values $E_k$.
We represent $\psi$ as $\psi(x,t) = \sum_n
c_{\eta, a, n}(t) \Psi_{a,n}(x)$, where $a = 1,2$, and  ${\hat A}$ has the above block - diagonal form in the basis of $\Psi_{a,n}$. These vectors are
normalized to unity ($\int\!\!d^3x\,{\Psi}_{an}^T \Psi_{an} = 1$). Further, we represent
\begin{equation}
Z= \int\!\! \,d c\,
{\rm exp}
\Big(-
\sum\limits\sb{\eta,n}T\,
{c}_{-\eta,n}^T [-i\eta+ E_n{\hat \beta}] c_{\eta,n}
\Big),
\end{equation}
where the system is considered
with the anti-periodic in time boundary conditions:
$\psi(t+T, x) =-\psi(t, x)$.
We use the decomposition
\begin{equation}\label{dec-0}
c_n(t) = \sum_{\eta=\frac{\pi}{T}(2k+1),\, k \in Z} e^{-i \eta t}
c_{\eta,n}.
\end{equation}
Integrating out the Grassmann variables $c_n$ we come to:
\begin{eqnarray}
\label{G_p_h_E}
&&
Z= \prod_{\eta>0}
%\eta=\frac{\pi}{T}(2k+1)
\prod_{n}
\bigl((\eta+E^{}_n)(-\eta+E^{}_n)T^2\bigr)=\prod_{\eta}
%\eta=\frac{\pi}{T}(2k+1)
\prod_{n}
\bigl((\eta+E^{}_n)T\bigr)
 =\prod\sb{n}\cos\frac{ T E^{\varphi}_n}{2},
\end{eqnarray}}

{The values $E_n$ depend on the parameters of the Hamiltonian,
with the index $n$ enumerating these values.
Eq. (\ref{G_p_h_E}) is derived as follows.
Recall that
in \eqref{dec-0}
the summation is over
$\eta=\frac{\pi}{T}(2k+1)$.
The product over $k$ can be calculated as in \cite{PhysRevD.12.2443}:
\begin{equation}
%\prod_{ k =-N}\sp{N }\Big(\big(\frac{\pi}{T}(2k+1) +  E^{\varphi}_n\big)T\Big)
%=
%\Bigl(
\prod_{k\in Z}
\Big(1 + \frac{ E_n T}{\pi(2k+1)}\Big)
%\Bigr)
%\Bigl(\prod_{k=-N}\sp{N} {\pi}{}(2k+1)\Bigr)
%\nonumber\\
%\to
%\approx
=
\cos\frac{E^{}_n T}{2},
%\prod_{k=-N+1}^N \pi (2k+1),
\label{DET0}
\end{equation}
Formally the partition function may be rewritten as
\begin{equation}
Z=  {\rm Det}^{1/2} \Bigl[ \partial_t + \hat A \Bigr]= \prod_n \cos\frac{E_n T}{2}
\end{equation}
The explanation that the square root of the determinant appears is that operator $\Bigl[ \partial_t + \hat A \Bigr]$ itself being discrcetized becomes the skew - symmetric matrix. Via the orthogonal transformations it may be made block - diagonal with the elementary $2\times 2$ blocks. In the latter form the functional integral is obviously equal to the square root of the determinant because for the 2 - component spinor $\eta$
\begin{equation}
\int d \eta \, {\rm exp}\Bigl[\eta^T \left(\begin{array}{cc}0 & -a \\a & 0 \end{array}\right) \eta\Bigr] = a = {\rm Det}^{1/2}\left(\begin{array}{cc}0 & -a \\a & 0 \end{array}\right)
\end{equation}
We get (see also \cite{PhysRevD.12.2443}):
\begin{eqnarray}
&&
\hskip -40pt
Z
=
\sum_{\{K_n\} = 0,1}
{\rm exp}
\Bigl(
 \frac{i T}{2} \sum_n
E_n-i T \sum_n K_n E_n
\Bigr)
 \label{G_p_h_E_0}
\end{eqnarray}}

Following \cite{PhysRevD.12.2443},
we interpret Eq. (\ref{G_p_h_E_0}) as follows. $K_n$
represents the number of occupied states with the energy $E_n$. These
numbers may be $0$ or $1$. The term $\sum_n E_n$ vanishes
if values $E_n$ come in pairs with the opposite signs (this occurs when the time reversal symmetry takes place).
We can rewrite the last expression in the form, when the integer numbers represent the numbers of occupied states of positive energy and the holes in the sea of occupied negative energy states:
\begin{eqnarray}
Z(T) = \sum_{\{K_n\} = 0,1} {\rm exp}\Bigl(  \frac{i T}{2} \sum_n |E_n|-i T \sum_n K_n |E_n| \Bigr)
 \label{G_p_h_E_}
\end{eqnarray}

After the Wick rotation we arrive at
\begin{eqnarray}
Z(-i/{\cal T}) = \sum_{\{K_n\} = 0,1} {\rm exp}\Bigl(  \frac{1}{2 {\cal T}} \sum_n |E_n|-\frac{1}{{\cal T}} \sum_n K_n |E_n| \Bigr),
 \label{G_p_h_E_2}
\end{eqnarray}
where $\cal T$ is temperature. This shows, that in equilibrium the configuration dominates with the vanishing numbers $K_n$. This corresponds to the situation, when all states with negative energy are occupied. This form of vacuum is intimately related with the anti - periodic in time boundary conditions imposed on $\psi$. The other boundary conditions would lead to the other prescription for the occupied states in vacuum.

%
%In the strong coupling limit  $\frac{\charge^2}{4\pi} \gg 1$
%we can evaluate the integral over $\varphi$
%in a stationary phase approximation.
%For the ordinary polarons this limit was considered
%originally by Landau and Pekar.
%
%In both limits mentioned above,
The values $E_n$ are given by
the solution of equation $({\hat A}^i_j\delta^a_b - \beta^a_b E_n) \zeta^b_i = 0$, where $\zeta^1, \zeta^2$ are the real - valued $n$ - component wave functions (i.e. $\zeta_i^a$ is the $2\times n$ matrix). Alternatively, we may solve equation
\begin{eqnarray}
0 &=& [\hat A + \partial_t] {\xi}\label{sch}
\end{eqnarray}
Here the real - valued $2\times n$ - component wave function $\xi$ is assumed to have the particular form
$\xi(x,t) = \zeta(x)e^{-\hat \beta E_n t} $.
Alternatively we may consider Eq. (\ref{sch}) as the equation for the pair of $n$ - component wave functions. Then the solutions of Eq. (\ref{sch}) are given by Eq. (\ref{TimeTranslationGeneral}).


\begin{thebibliography}{999}



\bibitem{Yang} C.N. Yang, "Thematic Melodies of Twentieth Century
Theoretical Physics : Quantization, Symmetry and Phase Factor",
in: International Conference on
Theoretical Physics, TH-2002, Paris, July 22-27, 2002,
D. Iagolnitzer, V. Rivasseau and J. Zinn-Justin eds., Birkh\"auser Verlag,
Basel-Boston-Berlin (2004), Ann. Henri Poincare {\bf 4}, Suppl. 2,   S9--S14
(2003).

\bibitem{Adler2013}
 S.L. Adler,
 Incorporating gravity into trace dynamics: the induced gravitational
action,
  arXiv:1306.0482

\bibitem{Horava2005}
P. Ho\v{r}ava,
Stability of Fermi surfaces and $K$-theory,
Phys. Rev. Lett. \textbf{95}, 016405 (2005).


\bibitem{VolovikBook} G.E. Volovik, {\it The Universe in a
Helium Droplet}, Clarendon Press,  Oxford (2003).

\bibitem{NeumannWigner}
J. von Neumann and E. Wigner,
 \"Uber merkw\"urdige diskrete Eigenwerte, Z. Phys. {\bf 30}, 465--467  (1929).

\bibitem{Novikov1981}
S. P. Novikov,
Bloch functions in a magnetic field and vector bundles. Typical dispersion relations and their quantum numbers,
 Dokl. Akad. Nauk SSSR {\bf 257}, 538--543 (1981).

\bibitem{Whitehead1942}
G.W. Whitehead,
Homotopy Properties of the Real Orthogonal Groups,
Annals of Mathematics, Second Series, Vol. 43, No. 1 (Jan., 1942), pp.
132--146.

\bibitem{StoneChiuRoy2011}
Michael Stone, Ching-Kai Chiu and Abhishek Roy,
Symmetries, dimensions and topological insulators: the mechanism behind the
face of the Bott clock,
J. Phys. A: Math. Theor. {\bf 44}, 045001 (2011).


\bibitem{CortesSmolin2013}
Marina Cortes and  Lee Smolin,
The Universe as a Process of Unique Events,
arXiv:1307.6167.


\bibitem{Z2013JHEP}
{``Gauge theory of Lorentz group as a source of the dynamical electroweak symmetry breaking''}, {}M.~A.~Zubkov, {}arXiv:1301.6971 [hep-lat], JHEP 1309 (2013) 044


\bibitem{latticebook}
I.Montvay, G.Munster, "Quantum fields on a lattice", Cambridge University Press, 1994

\bibitem{VZ2013gr}
M.~A.~Zubkov and G.~E.~Volovik,
``Emergent Horava gravity in graphene,''
  arXiv:1305.4665 [cond-mat.mes-hall], Annals of Physics, doi:10.1016/j.aop.2013.11.003;
{``Emergent gravity in graphene''},
  {\it Proceedings of the International Moscow Phenomenology Workshop
  (July 21, 2013 - July 25, 2013)},
  arXiv:1308.2249 [cond-mat.str-el]

\bibitem{PhysRevD.12.2443}
Dashen R~F, Hasslacher B and Neveu A 1975 {\em Phys. Rev. D\/} {\bf 12}(8)
  2443--2458

\end{thebibliography}
\end{document}